# Characterising Payload Entropy in Packet Flows

Baseline entropy analysis for network anomaly detection

Anthony Kenyon\*, Lipika Deka\*\*, David Elizondo\*\*

\*Hyperscalar Ltd, United Kingdom, tony.kenyon@ieee.org

\*\*Institute of Artificial Intelligence, School of Computer Science and Informatics, De Montfort University, United Kingdom

*Abstract*— Accurate and timely detection of cyber threats is critical to keeping our online economy and data safe. A key technique in early detection is the classification of unusual patterns of network behaviour, often hidden as low-frequency events within complex time-series packet flows. One of the ways in which such anomalies can be detected is to analyse the *information entropy* of the *payload* within individual packets, since changes in entropy can often indicate suspicious activity - such as whether session encryption has been compromised, or whether a plaintext channel has been co-opted as a covert channel. To decide whether activity is anomalous we need to compare real-time entropy values with baseline values, and while the analysis of entropy in packet data is not particularly new, to the best of our knowledge there are no published *baselines* for payload entropy across common network services. We offer two contributions: 1) We analyse several large packet datasets to establish baseline payload information entropy values for common network services, 2) We describe an efficient method for engineering entropy metrics when performing flow recovery from live or offline packet data, which can be expressed within feature subsets for subsequent analysis and machine learning applications.

## I. INTRODUCTION

Packet level *information entropy* can reveal useful insights into the types of content being transported across data networks, and whether that content type is *consistent* with the communication channels and service types being used. By comparing payload entropy with *baseline* values, we can ascertain - for example - whether security policy is being violated (e.g., an encrypted channel is being used covertly). To the best of our knowledge there are no published *baseline* information entropy values for common network services, and therefore no way to easily compare deviations from 'normal'. In this paper we analyse several large packet datasets to establish baseline entropy for a broad range of network services. We also describe an efficient method for recovering entropy during flow analysis on live or offline packet data, the results of which can be included as part of a broader feature subset, for subsequent analysis and machine learning applications.

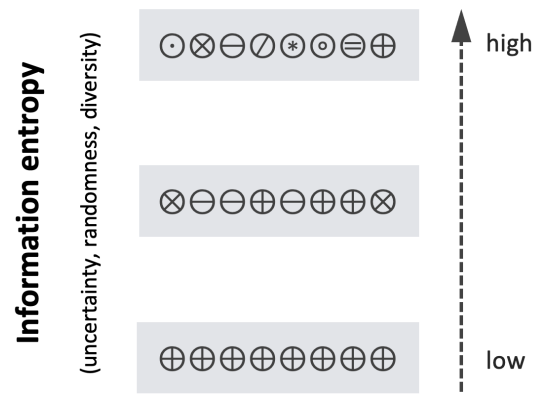

Fig. 1. Simplified illustration of information entropy for a fixed set of eight symbols. Lowest entropy is achieved with a monotonic set of repeating symbols (each with probability 1 of being predicted). Highest entropy is achieved when the full symbol set is used, with each symbol appearing randomly with equal probability.

### A. Background

Broadly, entropy is a measure of the state of *disorder*, *randomness*, or *uncertainty* in a system.[1] In the context of *information theory*, entropy was first described by Shannon in his seminal 1948 paper [SHAN48], which provides a mathematical framework to understand, measure, and optimise the transmission of information. Shannon formalised the concept of *information entropy* as a measure of the *uncertainty* associated with a set of possible outcomes: the higher the entropy, the more uncertain the outcomes.

---

[1] Definitions span multiple scientific fields, and the concept of 'order' can be somewhat subjective, however, for the purpose of this work we are concerned with *information entropy*. In terms of physics for example the configuration of the primordial universe has lowest overall entropy, since it is the most ordered and least likely state over a longer timeframe. We can consider entropy as the number of *configurations* of a system. It may also be viewed as a measure of the lack of information in a system.

Practically, if we consider entropy in data, we are interested in the frequency distribution of *symbols*, taken from a finite *symbol set*. The higher the entropy, the greater the diversity in symbols. Maximum entropy occurs when the symbolic content of data is unpredictable, so for example. if a file or network byte stream has high entropy, it follows that any symbol[2] is almost equally likely to appear next (i.e., the data sequence is unpredictable, close to random), see Fig. 1.

Shannon's entropy [SHAN48, SHAN15] - sometimes referred to as *information density*, is a measurement of the optimal average length of a message, given a finite symbol set (an *alphabet*). The use of entropy to measure uncertainty in a series of events or data is a widely accepted statistical practice in information theory [SHAN48, SHAN15].

We can compute the entropy of a discrete random event *x* using the following formula:

$$H(X) = -\sum_{i=1}^{n} p(x_i) \log_2(p(x_i))$$

Where H(x) is information entropy of a random variable *X*, with a finite set of *n* symbols. This formula is effectively the weighted average uncertainty for each outcome $x_i$, where each weight is the probability $p(x_i)$. P($x_i$) is the probability [0,1] of the occurrence of the $i^{th}$ outcome. Log base 2 is convenient to use since we are measuring entropy in bits (i.e., x ∈ {0,1}). The negative sign ensures non-negative entropy. In Section III we describe how we normalise entropy values to lie within the range 0 to 8 for the purpose of our packet analysis.

### B. Network packet and flow datasets

Machine learning is a powerful tool in cybersecurity, and particularly in its ability to detect anomalies, and cybersecurity researchers in network threat identification and intrusion detection are particularly interested in the analysis of large network packet and flow datasets. A survey of the composition of publicly available intrusion datasets is provided in [KEN20].

Packet datasets are typically large[3] high-dimensional, time-series data, often containing tens of millions of discrete packet events. Due to memory constraints and temporal complexity, it is common to abstract packet data into lower-dimensional containers called *flows*. Flows capture the essential details of packet streams in a compact extensible format, without the inherent complexity of raw packets [KEN20]. Packet flows offer a convenient lower dimension sample set with which to do cyber research.

A flow can be created using a simple tuple, to create a unique fingerprint with which to aggregate associated packets over time, based on the following attributes:

- Source and Destination IP Address
- Source and Destination Port identifier
- Protocol ID

Flows can be stateful[4], with additional logic and timeouts required to capture the full lifecycle of a flow. Flows may also be *directional* (i.e., unidirectional, or bidirectional), and flows may be *unicast*, *multicast*, or *broadcast* (one-to-one, one-to-group, one-to-all). While flows are essentially unique at any instant, they may not be unique across time - based solely on the tuple - since some attributes are likely to be reused in the distant future[5].

Modern public datasets used in network threat research often include high level flow summaries and metadata, but rarely include *payload content* in these summaries.[6] Since packet payload typically represents the largest contributor to packet size[7], it tends to be removed during the creation of flow datasets. Payload also adds complexity, in that it requires reassembly and in-memory state handling, across the lifecycle of a flow. Payload is often encrypted (see Fig 2), which means that many potentially useful featu.res are not accessible. There are also potential privacy and legal concerns, given that payload may contain confidential, sensitive, or personal information [KEN20]. For these reasons, we rarely see much information on packet payload within flow datasets and metadata summaries, other than simple volumetric metrics. As such, we have very few insights of what the actual content of the data being transferred looks like, at any point in time, and this lack of visibility can impair the detection of anomalous and suspicious activity that might exploit this feature. Specifically. the omission of such metrics in flow and packet data may inhibit detection of certain types of attack, and as discussed in Section II.

---

[2] Here we typically equate a byte to a symbol.

[3] Tens to hundreds of Gigabytes
[4] For example, TCP flows have a definite lifecycle, controlled by state flags
[5] For example, *port numbers* will eventually 'wrap around' once they reach a maximum bound - so large packet trace could contain two identical flows, but these are likely to be separated by a substantial time interval - unless there is a bug in the port allocation procedure.

C. *What kind of entropy metrics are useful network packet data*

In the context of network *packet data*, a variety of entropy measurements can be taken, and applied in the classification of network anomalies; based on the premise that deviation in entropy values from expected baselines can be indicators of specific threat vectors. For example, where synthetic attacks rely on simple script-based malware, features such as timing or address allocation may exhibit lower entropy (e.g., we might observe predictable packet intervals, payload sizes, port number allocations etc.)[8]. Naturally, skilled malware creators will attempt to mask such characteristics by reducing predictability (for example, by introducing randomised timing, more sophisticated address and port allocation techniques, perturbations in content etc.).

In practice, we can calculate entropy against several network features, including packet payload content, packet arrival times, IP addresses and service or port identifiers, as well as changes in entropy across time. Cybersecurity researchers have extended these concepts to a range of use cases in malware detection and content classification. For instance, techniques have been developed to identify anomalies in binary files, as well as encrypted network traffic, to indicate the presence of malicious code. We discuss several implementations of entropy in anomaly detection described in the literature in Section II.

Importantly, even where payload content is removed during flow creation, it is possible to extract useful information about payload composition, based on symbolic predictability. Metrics such as *information entropy* [SHAN48, SHAN15] can provide insights into the nature of encapsulated payload data and may be used as an indicator for security threats from covert channels, data exfiltration, and protocol compromise.

There is a subtle distinction here worth pointing out regarding the use of packet and flow-level entropy metrics:

- At the discrete *packet level*, individual packets arrive at typically random intervals[9], intermingled with many other packets, denoting different services and conversations. At a packet level the information entropy is effectively atomic, and we do not get a view of cumulative entropy over time, nor any changes in entropy over time.

- At a *flow level*, packets that are closely related over time are analysed statefully, and as a discrete group. For example, if a user sends an email there will typically be several related packets involved in the exchange, in two directions, and this collection of packets is termed a packet flow. Information entropy at this level can be useful in providing an overall perspective of the content of payload, and a per-packet perspective on any changes in entropy throughout the flow, by direction.

---

[6] Notably the UNB 2012 intrusion dataset did include some payload information encoded in Base 64,, however subsequent updates did not, due to the size implications.
[7] Typically, an order of magnitude larger than the protocol headers.
[8] This might be the case where malware contains simple data generating loops, and where events and data are allocated incrementally, at predictable intervals.

[9] On a large busy multi-protocol network with many active nodes we can reasonably assume this at least appears to be the case in practice.

Whilst there are times when individual packet entropy may be useful[10], ideally, we want to understand the cumulative entropy within packet payload, by direction, with the ability to identify any significant changes in entropy during the flow lifecycle.

*D. Information entropy baselines for network services*

Each packet traversing a network typically contains identifiers that associate that packet with a network service - for example, the File Transfer Protocol (FTP), where the payload of each packet usually represents a fragment of the content being transferred. Packet payload represents a rich source of high dimensional data, and techniques that examine this low-level information are termed *Deep Packet Inspection* (DPI). In the past we enjoyed almost complete visibility of this content since older network services (such as HTTP) encoded content in plaintext (i.e. unencrypted). Today, networks are dominated by *encrypted* services, such as HTTPS, where payload is effectively treated as a *'black-box'*[11], although, there remain some important legacy services that do not encrypt data, as shown in Fig. 2.

| Service | Port | Encrypted | Service | Port | Encrypted |
|---|---|---|---|---|---|
| ftp | 20,21 | n | ssh | 22 | y |
| telnet | 23 | n | kerberos | 88 | p |
| smtp | 25 | n | ldap | 389 | n |
| dns | 53 | n | ssl, https | 443 | y |
| bootp \| dhcp | 67,68 | n | smtps | 465 | y |
| tftp | 69 | n | nntps | 563 | y |
| http | 80 | n | ldaps | 636 | y |
| sftp | 115 | n | telnets | 992 | y |
| ntp | 123 | n | imaps | 993 | y |
| imap | 143 | n | ircs | 994 | y |
| netbios | 137-139 | n | pop3s | 995 | y |
| snmp | 161,162 | n | | | |

Fig. 2. Common well-known TCP and UDP 'well-known' ports for plaintext and cryptographic services. Here y=yes, n=no, and p=partial. Client applications that wish to use encrypted services typically start by exchanging cryptographic keys so that the rest of the conversation is secure. Note that some protocols use partially encrypted messaging, where typically the initial exchange is in plantext. These variations in use will be clearly reflected in payload entropy values.

We know that network services exhibit markedly different entropy profiles, since some are known to be plaintext, some partially or fully encrypted - as illustrated in Fig. 2. This gives us some intuition on what level of information entropy to expect when analysing the content of network traffic. However, since there are no published baselines for service level information entropy, even if we dynamically compute payload entropy (e.g., within an active flow) there are no 'ground truth' values with which to compare. Baseline data can prove very useful in determining whether the characteristics of flow content are deviating significantly from expected bounds, and this may be a strong indicator of anomalous activity, such as a covert channel, compromised protocol, or even data theft - as discussed in Section II.

*E. Information entropy expression in feature subsets*

Cybersecurity researchers using machine learning typically rely on small *feature subsets* with high predicted power to identify malicious behaviour. These features may be provided with a dataset or may be engineered from the dataset by the researcher. The composition and correlation strengths across these feature sets often vary, depending on the type of threat and the deployment context; hence the engineering of new features is an important area of research.

We described earlier that packet and flow datasets (particularly those publicly available [KEN20]) typically lack entropy metrics for payload content in their associated metadata[12]. In Section III we describe a methodology to enhance dataset flow metadata with information entropy features, and how we subsequently use that to calculate service level baseline information entropy values for various payload content types.

In the following section we describe related work in characterising various entropy metrics, and discuss examples where entropy has been used in anomaly detection and content classification to assist in network analysis and cybersecurity research.

---

[10] For example, where real-time intervention is critical.
[11] Without resorting to technologies that can unpack the data in transit (such as *SSL intercept*)

[12] For several reasons, though chiefly due payload being mainly encrypted nowadays, and the scale and resource challenges in decomposing and reassembling high dimensional content types.

## II. RELATED WORK

In the research it has been shown that *information gain* metrics using techniques such as entropy, can be useful in detecting anomalous activity. Encrypted traffic tends to exhibit a very different entropy profile to unencrypted traffic[13], specifically it tends to have much higher entropy values due to the induced unpredictability (randomness) of the data. Entropy has been widely used as a method to detect anomalous activity, and so is of interest in research such as intrusion detection, DDoS detection, and data exfiltration.

Early work by [GOUB06] characterises the entropy of several common network activities, as shown in Fig. 3. As discussed in [GOUB06], with standard cryptographic protocols (such as SSH, SSL, HTTPS) it is feasible to characterise which parts of traffic should have high entropy, after key exchange has taken place. Therefore, significant changes in entropy during a session may indicate malicious activity. During an OpenSSL or OpenSSH attack, entropy within an encrypted channel is likely to drop below expected levels as the session is perturbed; [GOUB06] suggests entropy scores would dip to approximately 6 bits per byte during such a compromise (i.e., entropy values of around 6, where we would normally expect it to by closer to 8).

| Data Source | Entropy (bits/byte) H^MLE | N HN |
|---|---|---|
| Binary executable (elf-i386) | 6.35 | 8.00 |
| Shell scripts | 5.54 | 8.00 |
| Terminal activity | 4.98 | 8.00 |
| 1 Gbyte e-mail | 6.12 | 8.00 |
| 1Kb X.509 certificate (PEM) | 5.81 | 7.80 ± 0.061 |
| 700b X.509 certificate (DER) | 6.89 | 7.70 ± 0.089 |
| 130b bind shellcode | 5.07 | 6.56 ± 0.24 |
| 38b standard shellcode | 4.78 | 5.10 ± 0.28 |
| 73b polymorphic shellcode | 5.69 | 5.92 ± 0.27 |
| Random 1 byte NOPs (i386) | 5.71 | 7.99 |

Fig. 3. Early analysis of entropy values from several content types, derived from [GOUB06]. As a point of reference Fig 4 and 5 provide more recent analysis of similar content types, where for example email has an average entropy ranging at between 5.40 (POP3) and 5.92 (SMTP).

In [LYD07] the authors use static analysis across large sample collections to detect compressed and encrypted malware, using entropy analysis to determine statistical variations in malware executables. In [GILB18] the authors use methods that exploit structural file entropy variations to classify malware content. in [HAN15] the authors use visual entropy graphs to identify distinct malware types. In [WANG11] the authors propose a classifier to differentiate traffic content types (including text, image, audio, video, compressed, encrypted, and Base64-encoded content) using Support Vector Machine (SVM) on byte sequence entropy values.

Analysis of the DARPA2000 dataset in [ZI10] lists the top 5 most important features as TCP SYN flag, destination port entropy, entropy of source port, UDP protocol occurrence and packet volume. [GOM12] describes how peer-to-peer Voice over IP (VOIP) sessions can be identified using entropy and speech codec properties with packet flows, based on payload lengths. In [ZEMP13] the authors use graphical methods for detecting anomalous traffic, based on entropy estimators.

In [GU05] the authors propose a propose an efficient behavioural-based anomaly detection technique, by comparing the maximum entropy of network traffic against a baseline distribution, using a sliding window technique with fixed time slots. The method is applied generically across TCP and UDP traffic and is limited to only three features (based on protocol information and destination port number). They are able detect fast or slow deviations in entropy, for example an increase in entropy during a SYN Flood. In [ROMAN08] the authors analyse entropy changes over time in PTR RR[14] DNS traffic to detect spam bot activity. In [ALTH15] the authors build on the concepts outlined in [GU05], capturing network packets and applying *relative entropy* with an adaptive filter to dynamically assess whether a change in traffic entropy is normal or contains anomaly. Here the authors employ several features, including *source and destination IP address*, *source and destination port*, and *number of bytes and packets sent and received*. [MAM16] describes an entropy based encrypted traffic classifier based on Shannon's entropy, and weighted entropy [CROLL13] and use of a Support Vector Machine (SVM).

---
[13] Unencrypted but *compressed* data may also show high entropy, depending on the compression algorithm and underlying data.

[14] Resource Records (RR) used to link IP addresses with domain names

In [ZHIY09] the authors propose a taxonomy for network covert channels based on entropy and channel properties, as well as suggesting prevention techniques. More recently, in [CHOW17] the authors focus on the detection of Covert Storage Channels (CSC) in TCP/IP traffic based on relative entropy of the TCP flags (i.e., deviation in entropy from baseline flag behaviour). In [HOM17] the authors describe entropy-based methods to predict the use of covert DNS tunnels, focussing on the detection of embedded protocols such as FTP and HTTP.

Cyber physical systems present a broad attack surface for adversaries [KEN18], and there there can be many active communication streams at any point in time. These channels can be blended into the victim's environment and used for reconnaissance activities and data exfiltration. In [LI22] the authors use of TCP payload entropy to detect real-time covert channels attacks on Cyber-Physical Systems (CPS). In [OZD22] the authors describe a flow analysis tool that provides application classification and intrusion detection, based on payload features that characterise network flows, including deriving probability distributions of packet payloads generated by *N*–gram analysis [DAM95].

Computing entropy in packet flows can be implemented by maintaining counters to keep track of the symbol distributions - however, this requires flow state to be maintained over time (we describe this further in Section III). This can be both computationally and memory intensive - particularly in large network backbones with many active endpoints. since flows may need to reside in memory for several minutes, possibly longer). For example, in [GU05] the authors state that their method requires constant memory and a computation time proportional to the traffic rate.

Where entropy is to be calculated in real time a different approach may be required. In [ARAC10], the authors offer a distributed approach to efficiently calculate conditional entropy to assist in detect anomalies in wireless network streams, by taking packet traces whilst an active threat in progress. They propose a model based on the Hierarchical Sample Sketching (HSS) algorithm, looking at three features of the IEEE 802.11 header: frame length, duration/ID, source MAC address (final 2-bytes) to compute conditional entropy.

III. METHODOLOGY

In this section we discuss the methodology used to calculate both baseline values and the individual flow level entropy feature values.

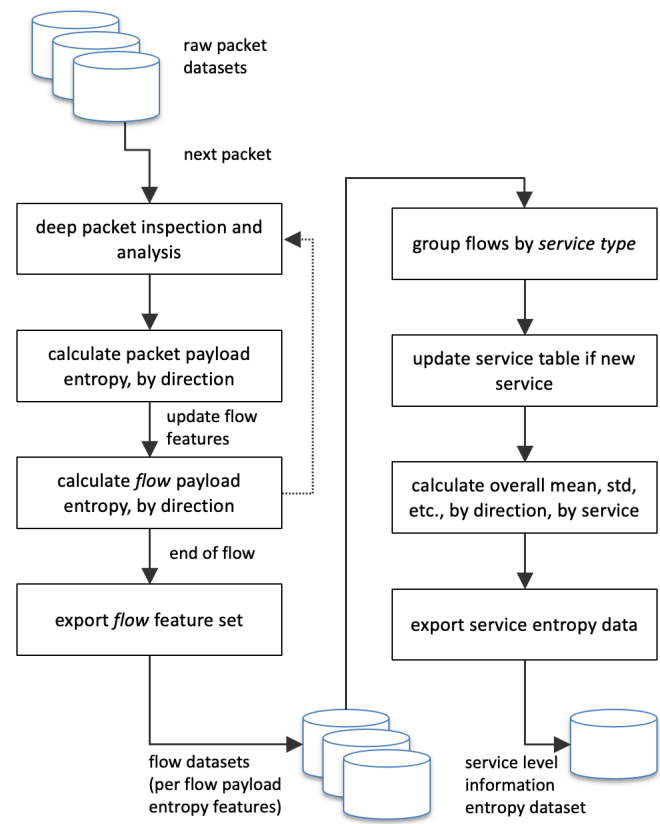

Fig. 4. Two phase analysis for calculating service baseline metrics for payload entropy. Packets are first grouped into logical flows to ensure that we are tracking entropy changes for each discrete flow duration. All flow entropy values are then grouped by service types and overall basleine entropy metrics calculated. Note that the contribution of each dataset is weighted by sample size (to avoid the case where a smaller anomalous dataset distort the overall metrics)[15].

A. *Baseline data processing methods*

The methodology we used to analysis information entropy in packet traces and calculate service level baselines can be summarised in two phases, as illustrated in Fig. 4. The first phase analyses a set of raw packet datasets (listed in Fig. 5), calculating payload entropy per packet, grouping packets into flows, and calculating the final payload entropy per flow. The second phase takes the resulting flow datasets, grouping all flows by *service type* (i.e., based on TCP or UDP destination port), and calculates the service level baseline entropy features, for all datasets.

---
[15] We also ignore samples that are clearly labelled as anomalous in datasets such as those used in intrusion detection, since these samples may include values outside the expected baseline range.

Sample size is included in the analysis, since some services are more widely represented in the packet distributions than others (for example in a typical enterprise network packet trace we would expect to see a high percentage of web traffic, and much less SSH traffic [KEN20]).

### B. Data sources

As part of this research detailed analysis was performed to characterise payload entropy values, for a range of *well-known* and *registered* services, averaged across a range of environments, as shown in Fig. 4. Raw packet data were sourced from several widely used public sources (described in [KEN20]) as well as recent live capture traces. Raw PCAP files were converted into flow records, with payload entropy reconstructed for common TCP and UDP protocols.

| Dataset | Year | Packets | Samples | Publisher |
|---|---|---|---|---|
| UNB2017 | 2017 | 56,370,702 | 51,842,003 | UNB |
| File: UNB 2017 Monday-Friday workinghours (5 files)) | | | | |
| UNSW-NB15 | 2015 | 4,592,899 | 1,834,238 | UNSW |
| Files: 17-2-15/1, 17-2-15/27 (2 files) | | | | |
| NETRESEC | 2017 | 2,274,747 | 216,925 | NETRESEC |
| File: 4SICS-GeekLounge-151022 | | | | |
| CSE-CIC-IDS2018 | 2018 | 473,278 | 205,950 | UNB |
| File: capEC2AMAZ-O4EL3NG-172.31.69.19 | | | | |
| LIVE_TRACE | 2023 | 355,818 | 127,026 | Live |
| File: LT-2023-06-16-browser-vid | | | | |
| UNB2012 | 2012 | 8,177 | 6,357 | UNB |
| File: 582 flows extracted | | | | |
| TOW-IDS | 2020 | 791,611 | 100 | IEEE |
| File: Automotive_Ethernet_with_Attack_original_10_17_20_04_test | | | | |
| TOTAL | | 64,867,232 | 54,232,599 | |

Fig. 5. Datasets used in entropy calculations. The majority of samples were taken from the full UNB 2017 dataset (containing over 56 million packets), although several other datasets were tested to assess consisttency. These datasets are documented in [KEN20]. The original flow summaries provided with some of these sources were not used, since they lacked essential payload features, and in some issues with the original floe recovery. Therefore we reconstruct all flows and exposed aditional entropy metadata. In the table 'samples' indicates observations that matched a specific service type. Note that by 'sample' we mean the number of actual packets used in the analysis, given that network packet traces may contain packets that are either in error or not relevant to analysis.

In total over 54 million packets were sampled. Datasets were selected to avoid sources with known large distortions to ensure that values were statistically consistent across datasets (and where possible, labelled anomalies are excluded). Results are also weighted, per service type, with respect to sample size, so that the contribution of each packet trace is proportionate (i.e., small dataset samples containing outliers do not distort overall results). Where anomalies are labelled, we exclude these labelled events from the calculations - therefore the estimates are for known 'normal' traffic.

### C. Flow information entropy calculation methods

In our implementation[16] we provide measures for characterising mean payload entropy in *both* flow directions (inbounds and outbound), as part of the feature engineering process. Since payload is typically fragmented over multiple packets, entropy may vary during the lifespan of the flow, and will be summarised from multiple consecutive samples[17]. Our implementation is based on a modified Krichevsky-Trofimov (KT) estimator [KRICH81], which estimates the probability of each symbol in a particular symbol alphabet.

- A KT class is implemented as a bi-directional in-memory cache (a hash table of symbol frequencies, of capacity 256 - since we are dealing with a byte encoded stream[18]) and holds symbolic frequency data for the two payloads.

- The flow tuple is used to index the cache. Each flow effectively has a single cache entry, with statistics and state tracked for both flow directions.

- Cache entries are updated when each new packet is encountered. The updates are assigned to the associated flow record in the cache, or a new flow record is created, and those updates applied.

- Cumulative flow entropy values are recalibrated per packet, per direction, based on the current payload symbol frequencies and the running payload length.

---

[16] These features have been implemented in the GSYNX analysis suite, which will be made available at [GIT].
[17] For TCP these content 'fragments' may represent encapsulated data and/or other protocol plus data.
[18] Each symbol is 8 bits wide, corresponding to 0-255

- Once a flow is finalised, the final payload information entropy values are calculated using the total length of the payload, against the cumulative symbol frequencies, per direction.

Even if a flow is not terminated correctly, a cumulative entropy value is maintained and exported during flow dataset creation. The final entropy value will be in the range 0 to 8, for reasons described below.

### D. Applying Shannon's entropy to byte content

When we apply Shannon's method (described in Section I) to text content, we assume that symbols may be encoded in 8-bit bytes[19]. Since each bit has two possible values (0 or 1), the total number of possible combinations for a byte is $2^8$, or 256. This gives us a range of entropy values between 0 (low) and 8 (high).

- Where only one symbol is repeated, the symbol has a probability of 1, and hence the formula is resolved to:

$$H = -\log_2(1) = 0$$

- Where all symbols are used, each symbol has a probability of 1/256, and hence the formula is resolved to:

$$H = -\sum_{i=1}^{256} \left(\frac{1}{256}\right) \log_2 \left(\frac{1}{256}\right)$$

$$= -256 \left(\frac{1}{256}\right) \log_2 \left(\frac{1}{256}\right)$$

$$= -\log_2 \left(\frac{1}{256}\right)$$

$$= 8$$

Giving a low to high range of entropy values from 0 to 8, which is the range we apply in our analysis.

### E. Flow recovery methods

Our analysis required the use of specially developed software called GSX [KEN23] to perform flow recovery from large packet traces in *pcap* [PCAP] format. GSX performs advanced flow recovery, including stateful reconstruction of TCP sessions, with additional feature engineering to calculate a broad variety of metrics, including features characterising payload[20]. Payload entropy was calculated in both directions (outbound and inbound, with respect to the packet flow source[21]).

Flow collection is also possible in real-time, using common network tools and hardware [HOFS14, PATTS12], using industry standard and extensible schemas, as provided with network flow standards such as IPFIX [RFC7011], and later versions of NetFlow. Depending on available resources, in-flight capture may differ, for example by employing sampling and sliding window methods [CHEN14].

### F. Implementation challenges

This section highlights a number of challenges in efficient entropy calculation within the flow recovery process:

**Language Sensitivity**: Large packet traces may hold millions of packet events (see Fig. 5), resulting in hundreds and thousands of flows [KEN20]. Flow recovery is both memory and computationally expensive, and the time to produce an accurate flow dataset with a broad list of useful feature set (e.g., 100 features) may take hours, depending on the implementation language, and efficiency of the design. For example, an interpreted language such as python is likely to be an order of magnitude slower when compared with languages like Go, C, C++ or Rust.

**Cache Size**: Large packet traces may hold millions of packet events (see Fig. 5), resulting in hundreds and thousands of flows [KEN20]. Flow recovery is both memory and computationally expensive and is sensitive to the composition of the packet data and length of the trace. For example, with a short duration trace from a busy Internet backbone there may be tens of thousands of flows that never terminate within the lifetime of the trace - in which case all these flows will need to be maintained in cache memory until the last packet is processed. Conversely, a longer packet trace from a typical enterprise network may contain many thousands of flows that terminate across the lifetime of the packet trace, and so the cache size will tend to grow to reach a steady state and then gradually shrink.

---

[19] For example, with the ASCII character set. Unicode text may be encoded in 8, 16 or 32-bit blocks.

[20] In this analysis we focus primarily on TCP and UDP protocols
[21] Outbound meaning that the initiator of the flow is sending data to a remote entity. Here source can be thought of as the end-point IP address.

**Entropy Calculation**: to avoid multiple processing passes on the entire packet data, flow level information entropy analysis can be implemented within the flow recovery logic. As described earlier, by using the flow tuple as an index to a bidirectional hash table, symbol counts can be updated efficiently on a per packet basis. Each symbol type acts as a unique key to a current counter value. Changes in entropy are therefore detectable within the lifespan of the flow, by direction. We include additional measures of entropy variance, by flow direction, which can be another useful indicator for major entropy deviations from baselines.

**Real-time flow recovery**: Flows can be recovered from offline packet datasets and archives. They can also be assembled in real time, using industry standards such as NetFlow [KERR01, RFC3954] and IPFIX [RFC6313, RFC7011, RFC7012]. Since our primary interest is in recovering these features from well-known research datasets, we do not implement real-time recovery of payload entropy from live packet captures. Prototyping however suggests that using our implementation in a compiled language with controlled memory management (such as Rust or C++) is practical[22]. It is also possible to avoid some of the time and memory complexity of flow recovery if we only want to record flow level payload entropy from live packet data, by using simpler data structures, although protocol state handling is still required [KEN23].

IV. ANALYSIS

In this section we present our analysis on expected baseline entropy values across a broad range of common network services, together with our findings and some notes on applications.

*A. Baseline payload information entropy*

The results of our analysis are shown in Fig. 6. This table shows average entropy values for a common network services, together with their overall mean, together with directional mean and standard deviations[23]. Note that the service list has been derived dynamically from the datasets cited in Section III.B. For further information on specific port allocations and service definitions see [WIKIP23].

The table forms a consistent view of expected 'ground truth' across a range of deployment contexts. As might be expected, services that are encrypted (such as SSH) tend to exhibit high entropy (close to 8.0), whereas plaintext services (such as DNS) tend have low to mid-range entropy values. It is worth noting that entropy values close to zero are unlikely to be observed in real-world network traffic, since this would equate to embedding symbol sequences with very little variety[24], and even plaintext messages are likely to have entropy values in the range 3-4. This may not be immediately obvious and so in Fig. 8 we show how low entropy values (close to zero) might be achieved by severely restricting the symbolic content artificially.

As noted earlier, the sample count indicates the number of packet level observations found in the data, and here we see wide variation in the distribution frequency across services represented. For example, web based flows (HTTPS and HTTP) dominate the dataset composition, whereas older protocols such as TFTP) are less well represented. Whilst sample size can be used as a rough analogue for confidence in these baseline estimates, we have excluded services that had extremely low representation.

We observed strong consistency across many of the packet traces, however it is worth noting that in practice some services may exhibit deviation in entropy from baselines during normal activities, and this may depend on the context. For example, some services are specifically designed to encapsulate different types of file and media content that could vary markedly in composition and encoding (e.g., compressed video and audio content will tend to exhibit high entropy, whereas uncompressed files may exhibit medium-range entropy).

---

[22] Real-time performance (without packet drop) will depend to some extent on how many other features are to be calculated alongside entropy, and the complexity of those calculations.

[23] In a small number of cases only one flow direction is recorded, typically because such protocols are unidirectional or broadcast in nature.

[24] For example, by sending a block of text containing only repetitions of the symbol 'x'.

Also note that peer-to-peer protocols may also be encapsulated within protocols such as HTTP and HTTPS, and this can make it harder to characterise the true underlying properties of the content (without deeper payload inspection) as highlighted in [WANG11][25]. It is therefore important to use appropriate domain expertise when performing analysis, with an understanding of the communication context.

| service | port | mean 2-way | std | mean out | mean in | std out | std in | samples |
|---|---|---|---|---|---|---|---|---|
| ssh | 22 | **7.631** | 0.650 | 7.568 | 7.695 | 0.605 | 0.694 | 678,138 |
| cisco.ssm | 465 | **7.587** | 0.731 | 7.555 | 7.617 | 0.643 | 0.815 | 80,032 |
| openflow | 6653 | **7.576** | 0.374 | 7.544 | 7.607 | 0.363 | 0.385 | 370 |
| https | 443 | **7.517** | 1.314 | 7.426 | 7.608 | 1.233 | 1.396 | 19,723,054 |
| mftp | 5402 | **7.189** | 1.589 | 7.208 | 7.170 | 1.825 | 1.353 | 193 |
| kerberos | 88 | **7.067** | 1.112 | 6.987 | 7.145 | 1.070 | 1.148 | 40,505 |
| radius.coa | 3799 | **7.000** | 1.360 | 6.235 | 7.765 | 2.183 | 0.537 | 357 |
| stun | 5349 | **6.952** | 2.590 | 6.800 | 7.104 | 2.601 | 2.579 | 492 |
| radius.acc | 1813 | **6.885** | 0.390 | 6.558 | 7.189 | 0.627 | 0.153 | 329 |
| bitcoin.rpc | 8332 | **6.626** | 1.883 | 5.902 | 7.351 | 1.919 | 1.847 | 235 |
| openvpn | 1194 | **6.599** | 1.084 | 5.811 | 7.385 | 1.406 | 0.762 | 1,465 |
| rtcp | 5005 | **6.584** | 1.720 | 6.356 | 6.812 | 2.011 | 1.430 | 195 |
| cms | 5318 | **6.523** | 1.790 | 5.867 | 7.179 | 2.099 | 1.467 | 285 |
| radsec | 2083 | **6.497** | 1.230 | 6.550 | 6.444 | 1.154 | 1.306 | 179 |
| netconf.tls | 6513 | **6.414** | 2.088 | 6.027 | 6.800 | 2.166 | 2.011 | 1,151 |
| ldap | 389 | **6.308** | 1.665 | 6.467 | 6.148 | 1.458 | 1.870 | 138,068 |
| nessus | 1241 | **6.230** | 1.091 | 5.836 | 6.625 | 0.886 | 1.296 | 724 |
| activedir | 445 | **6.227** | 1.609 | 6.312 | 6.140 | 1.514 | 1.700 | 115,371 |
| ipfix | 4739 | **5.974** | 1.928 | 6.710 | 5.237 | 1.547 | 2.309 | 254 |
| smtp | 25 | **5.913** | 0.096 | 6.056 | 5.769 | 0.090 | 0.101 | 235,267 |
| imap | 143 | **5.893** | 0.231 | 5.697 | 6.090 | 0.249 | 0.212 | 304,522 |
| l2f | 1701 | **5.888** | 1.484 | 5.688 | 5.885 | 1.533 | 1.435 | 419 |
| vxlan | 4789 | **5.805** | 1.726 | 5.737 | 2.173 | 1.943 | 1.509 | 320 |
| netbios.ds | 138 | **5.775** | 0.181 | 5.775 | 0.000 | 0.180 | 0.001 | 25,544 |
| afp | 548 | **5.771** | 0.782 | 5.761 | 0.009 | 0.000 | 0.782 | 9,791 |
| diameter | 3868 | **5.685** | 1.783 | 5.354 | 6.016 | 1.794 | 1.771 | 235 |
| ident.reg | 4604 | **5.545** | 1.672 | 5.137 | 5.952 | 1.822 | 1.523 | 226 |
| mswins | 1512 | **5.332** | 1.967 | 5.008 | 5.657 | 1.867 | 2.066 | 262 |
| pop3 | 110 | **5.246** | 0.417 | 4.666 | 5.814 | 0.511 | 0.290 | 29,013 |
| ntp | 123 | **5.088** | 0.458 | 5.166 | 5.009 | 0.318 | 0.598 | 73,868 |
| ftp.data | 21 | **4.908** | 0.410 | 5.110 | 4.706 | 0.414 | 0.406 | 255,303 |
| http | 80 | **4.899** | 2.813 | 4.441 | 5.358 | 2.601 | 3.024 | 28,573,190 |
| kubernetes.apisvr | 6443 | **4.785** | 2.885 | 4.515 | 5.056 | 2.812 | 2.958 | 143 |
| dns | 53 | **4.738** | 0.399 | 4.295 | 5.180 | 0.364 | 0.435 | 3,637,086 |
| ms.kms | 1688 | **4.697** | 0.613 | 3.991 | 5.403 | 0.614 | 0.611 | 1,921 |
| cifs | 1293 | **4.552** | 4.199 | 4.319 | 4.772 | 4.085 | 4.312 | 559 |
| tftp | 69 | **4.262** | 1.311 | 4.262 | 0.000 | 1.311 | 0.000 | 62 |
| bootp.server | 67 | **3.942** | 0.722 | 4.030 | 1.629 | 0.875 | 0.282 | 144 |
| bootp.client | 68 | **3.868** | 0.710 | 3.953 | 1.623 | 0.864 | 0.284 | 144 |
| socks.proxy | 1080 | **3.849** | 2.124 | 3.865 | 3.834 | 1.016 | 3.231 | 1,865 |
| rip | 520 | **3.836** | 0.287 | 3.836 | 0.000 | 0.287 | 0.000 | 1,258 |
| rtp | 5004 | **3.370** | 1.030 | 4.045 | 2.694 | 0.862 | 1.198 | 1,338 |
| hsrp | 1985 | **3.243** | 1.398 | 3.169 | 3.316 | 1.574 | 1.222 | 521 |
| netbios.ss | 139 | **3.151** | 2.882 | 2.913 | 3.388 | 3.009 | 2.751 | 70,994 |
| http.2 | 8080 | **3.129** | 0.977 | 3.092 | 3.158 | 0.851 | 1.082 | 14,828 |
| trpwire | 9898 | **3.060** | 1.412 | 2.755 | 3.207 | 0.590 | 1.648 | 631 |
| bgp | 179 | **3.039** | 1.208 | 5.842 | 0.232 | 1.464 | 0.943 | 27,887 |
| ipsec.nattrv | 4500 | **3.003** | 4.115 | 3.000 | 3.000 | 4.113 | 4.117 | 16,501 |
| kerberos.rsh | 514 | **2.621** | 0.736 | 2.540 | 0.346 | 0.662 | 0.731 | 999 |
| telnet | 23 | **2.518** | 2.138 | 2.094 | 2.943 | 1.882 | 2.395 | 17,114 |
| https.proxy.1 | 4444 | **2.473** | 1.463 | 2.509 | 2.335 | 0.557 | 2.065 | 897 |
| ms.dcom | 1029 | **2.340** | 1.396 | 2.244 | 2.340 | 1.219 | 1.304 | 1,032 |
| pptp | 1723 | **2.315** | 1.671 | 3.535 | 1.095 | 1.468 | 1.873 | 2,724 |
| https.proxy.2 | 4445 | **2.302** | 1.105 | 2.023 | 2.477 | 0.784 | 1.077 | 782 |
| ms.sql.svr | 1433 | **2.253** | 0.907 | 2.268 | 2.238 | 0.894 | 0.905 | 905 |
| ms.mq | 1801 | **2.233** | 1.505 | 1.861 | 2.604 | 1.796 | 1.214 | 875 |
| netbios.ns | 137 | **2.225** | 1.475 | 3.583 | 0.856 | 1.188 | 1.762 | 114,395 |
| citrix | 1494 | **2.078** | 1.015 | 2.039 | 2.117 | 1.077 | 0.952 | 854 |
| radius.auth | 1812 | **2.008** | 1.505 | 2.412 | 1.589 | 0.984 | 2.027 | 963 |
| sip.tls | 5061 | **1.829** | 1.675 | 1.817 | 1.750 | 1.003 | 2.072 | 863 |
| ms.sql.mon | 1434 | **1.796** | 1.235 | 1.743 | 1.842 | 1.207 | 1.262 | 896 |
| mysql | 3306 | **1.761** | 1.204 | 1.830 | 1.595 | 0.821 | 1.273 | 1,047 |
| postgresql | 5432 | **1.604** | 1.086 | 1.411 | 1.552 | 0.731 | 1.071 | 702 |
| sip | 5060 | **1.587** | 1.195 | 2.348 | 0.709 | 0.863 | 1.502 | 1,270 |
| lpd | 515 | **1.251** | 0.688 | 1.046 | 0.206 | 0.094 | 0.593 | 413 |
| rtmp | 1935 | **1.197** | 1.187 | 1.209 | 1.184 | 1.016 | 1.358 | 734 |
| nfs | 2049 | **1.121** | 1.322 | 1.018 | 1.051 | 0.875 | 1.421 | 763 |
| ms.mms | 1755 | **1.117** | 1.271 | 1.104 | 1.064 | 1.071 | 1.291 | 1,414 |
| torpark | 81 | **0.921** | 1.549 | 0.873 | 0.968 | 1.658 | 1.440 | 1,071 |
| http.1 | 8008 | **0.914** | 1.269 | 0.751 | 0.955 | 0.852 | 1.315 | 663 |
| bitcoin | 8333 | **0.896** | 1.282 | 0.742 | 0.931 | 0.913 | 1.297 | 691 |
| irc | 194 | **0.891** | 1.815 | 1.279 | 0.503 | 2.299 | 1.331 | 23 |
| imaps | 993 | **0.454** | 1.340 | 0.629 | 0.278 | 1.906 | 0.774 | 527 |
| rlogin | 513 | **0.331** | 0.501 | 0.211 | 0.258 | 0.250 | 0.752 | 347 |
| isakmp | 500 | **0.317** | 0.767 | 0.182 | 0.308 | 0.582 | 0.902 | 401 |
| ftps.ctrl | 990 | **0.303** | 0.792 | 0.000 | 0.303 | 0.000 | 0.792 | 316 |
| ldaps | 636 | **0.297** | 0.786 | 0.000 | 0.297 | 0.000 | 0.786 | 360 |
| bgmp | 264 | **0.282** | 0.770 | 0.000 | 0.282 | 0.000 | 0.770 | 335 |
| kerberos.adm | 749 | **0.270** | 0.754 | 0.000 | 0.270 | 0.000 | 0.754 | 321 |
| msxchange.rout | 691 | **0.270** | 0.755 | 0.000 | 0.270 | 0.000 | 0.755 | 384 |
| kerberos.login | 543 | **0.269** | 0.756 | 0.000 | 0.269 | 0.000 | 0.756 | 312 |
| pop3s | 995 | **0.267** | 0.754 | 0.000 | 0.267 | 0.000 | 0.754 | 417 |
| nntps | 563 | **0.266** | 0.751 | 0.000 | 0.266 | 0.000 | 0.751 | 350 |
| smux | 199 | **0.264** | 0.746 | 0.000 | 0.264 | 0.000 | 0.746 | 449 |
| kerberos.pwd | 464 | **0.263** | 0.748 | 0.000 | 0.263 | 0.000 | 0.748 | 317 |
| doom | 666 | **0.261** | 0.744 | 0.000 | 0.261 | 0.000 | 0.744 | 359 |
| smtp.ms | 587 | **0.259** | 0.745 | 0.000 | 0.259 | 0.000 | 0.745 | 408 |
| snmp | 161 | **0.257** | 0.709 | 0.125 | 0.314 | 0.513 | 0.901 | 919 |
| auth | 113 | **0.254** | 0.726 | 0.000 | 0.254 | 0.000 | 0.726 | 424 |
| whois | 43 | **0.243** | 0.724 | 0.000 | 0.243 | 0.000 | 0.724 | 326 |
| cisco.tdp | 711 | **0.241** | 0.718 | 0.000 | 0.241 | 0.000 | 0.718 | 336 |
| ftp.cmd | 20 | **0.238** | 0.713 | 0.000 | 0.238 | 0.000 | 0.713 | 376 |
| macos.server | 106 | **0.232** | 0.707 | 0.000 | 0.232 | 0.000 | 0.707 | 377 |
| ripng | 521 | **0.215** | 0.730 | 0.000 | 0.215 | 0.000 | 0.730 | 41 |
| TOTAL | | | | | | | | 54,225,733 |

Fig. 6. Mean and standard deviation for payload entropy values averaged over multiple traffic sources, by flow direction (outbound and inbound, with respect to session initiation). Note that encrypted services such as SSH, SSL and HTTPS have average entropy values closer to 8.0, whereas unencrypted services such as Telnet, LDAP and NetBios have low entropy values, indicating that the payload has a larger proportion of plaintext data. This data was aggregated across mutiple deployment conexts (enterprise, network backbone, industrial etc.). To account for the wide variations in sample sizes for specific protocols between packet traces, we weight the means by sample size, so that potential outliers in small packet traces do not influence the overall mean results disproportionately.

---

[25] It is also worth noting that system administrators sometimes change the port allocations to mask service usage or conform to strict firewall rules (this is not uncommon practice with protocols such as FTP and OpenVPN for example).

Note that the *standard deviation* metrics are also presented in Fig. 6, on a per service, per flow direction basis. For most of services we analysed the standard deviation sits typically below 2.0. Higher variance is more likely to be found in services that are used to encapsulate and transport a variety of content types, particularly where a service is normally unencrypted (e.g., web-based protocols such as HTTP, and file transfer protocols such as CIFS). This higher variance is likely to be attributable to the wide variety of content types encapsulated (some of which might be encrypted or compressed at source).

B. *Interpreting entropy variations*

Since baseline information entropy values are generally consistent across a range of deployment contexts, a *deviation* in entropy may be useful to indicate the type of content being transferred. and whether this is normal or anomalous behaviour. For example, if a user is uploading an encrypted file using FTP we would anticipate a higher entropy value than the expected baseline (around 4.1) for a particular flow. If this transfer were to an unknown external destination, then this might raise suspicion about the possibility of data exfiltration. Here again some domain expertise can be valuable, coupled with local knowledge on user behaviour and the type of data being moved.[26]

Embedded malware and executable files, often compressed, may also be an indicator of unusual content. For reference, several common types of file content have been analysed and their respected entropy values given in Fig. 7, together with their post-encrypted entropies. Note that compressed content[27] exhibits entropy close to 8, as we might expect, due to symbol repetition compaction. In these tests AES encryption was used with 256-byte keys, although other key sizes yielded similar entropy results). Domain expertise may be useful in determining whether a flow with very high entropy is likely to be encrypted or compressed - for example by examining a flow to establish whether entropy is consistent throughout its lifespan.

---

[26] For example, moving encrypted data over a plaintext channel such as FTP to an external competitor could be suspicious.
[27] Here we tested ZIP compression, although other comparable compression methods will present similar entropy results. The more efficient the compression method the higher the entropy.

| Description | Type | Plaintext entropy | File Size bytes | AES 256 entropy |
|---|---|---|---|---|
| wins executable | EXE | 5.8091 | 10842680 | 7.9950 |
| packet trace | PCAP | 7.9581 | 26073074 | 7.9999 |
| code repo python | ZIP | 7.9991 | 1253078 | 7.9999 |
| book - the illiad homer | ZIP | 7.9976 | 198358 | 7.9991 |
| audio_recoding.wav | ZIP | 7.9975 | 2440233 | 7.9999 |
| book: the illiad homer | PDF | 7.9963 | 33688009 | 8.0000 |
| research paper | PDF | 7.9578 | 303396 | 7.9993 |
| tuitorial | PDF | 6.4578 | 832414 | 7.7096 |
| wiki page | PDF | 5.7842 | 5869349 | 7.9955 |
| book: the illiad homer | TXT | 4.6104 | 525670 | 7.9996 |
| text file summary table | TXT | 2.9057 | 3308 | 7.7000 |
| spreadsheet | XLSX | 3.4905 | 1028 | 7.8237 |
| AirQualityUCI | CSV | 3.4489 | 629862 | 7.9995 |
| hires image new york colour | PNG | 7.9934 | 11921184 | 8.0000 |
| hires image new york bw | PNG | 7.9013 | 5701586 | 8.0000 |
| hires image new york colour | JPEG | 7.9895 | 3959743 | 7.9999 |
| table image bw | JPEG | 7.8975 | 651932 | 7.9994 |
| cheat sheet colour | JPEG | 7.8364 | 1261687 | 7.9998 |
| web asset icon grey | PNG | 7.1974 | 565 | 7.6674 |
| hires image new york colour | TIFF | 6.6273 | 21575314 | 7.9998 |
| audio_recoding.wav | WAV | 7.2539 | 2828586 | 7.9995 |

Fig. 7. Common file types and entropy values. 'Plaintext' here means unencrypted. On the right we also see corresponding entropies for AES 256 encrypted files. We use just the 256 block size as illustrative, since larger block size does not significantly improve the results - given these are close to 8.0 already. Note that zip compressed files and encrypted files tend to have entropy close to 8.

To illustrate the relationship between entropy and symbolic variety more clearly, Fig. 7 illustrates entropy values for three test files, plus an example of a well-known English text.

| Description | Type | Plaintext entropy | File Size bytes | AES 256 entropy |
|---|---|---|---|---|
| book: a midsummer nights dream | TXT | 4.8417 | 120868 | 7.9987 |
| symbol_test_full | TXT | 6.5850 | 96 | 6.3542 |
| symbol_test_duo | TXT | 1.0358 | 1623 | 5.6022 |
| symbol_test_mono | TXT | 0.0075 | 1617 | 4.0235 |

Fig. 8. Illustrates the effects of symbolic content on entropy values using four raw text files. The three special **'symbol_test'** files have limited symbolic alphabets. **symbol_test_mono** comprises only 1 repeated symbol, with corresponding entropy close to zero. **symbol_test_duo** contains two repeated symbols, with corresponding entropy close to 1. **symbol_test_full** contains a richer alphabet of 96 symbols (A-Z, a-z, plus punctiation etc.), with corresponding entropy rising above 6. The final example is a text representation of a book, which has lower entropy than **symbol_test_full** because of the frequent symbol repetitions typical in written language (some letters and sequeqnces are far more common than others). Encrypted versions of these files also exibit wide entropy variations. lower values due to the lack of symbol variety in the source data.

The test files were constructed with increasing levels of symbolic variety, and we can clearly see corresponding changes in entropy. From this, and the examples in Fig. 7, we can reasonably infer that typical written messages and content would be expected to have entropy in the mid-range (between 3 and 5).

### C. Applications

As mentioned earlier, it is possible to detect threats, even with encrypted traffic streams, where entropy deviates significantly from expected ranges, or changes during the lifespan of the session. Where content is being passed over a network, high entropy tends to indicate that data is either encrypted or compressed[28]. Knowing this we can analyse *payload entropy* dynamically and use this as an indicator for encrypted data streams, potentially identifying covert channels [LAMP73, ZAND07] and encapsulated malware. For example:

- For example, where a particular service is expected to encode content as plaintext (such as DNS), the detection of high entropy may indicate the presence of a covert channel, which could be used for *data exfiltration*.

- Unexpected plaintext on an encrypted channel may indicate a misconfiguration of the SSL/TLS encryption settings, or a security vulnerability in the system. For example, the Heartbleed vulnerability found in OpenSSL in 2014 is triggered when malicious heartbeat message causes the SSL server to dump plaintext memory contents across the channel [HBCVE].

- On an encrypted channel (such as an SSH tunnel or an HTTPS session), after a connection is established (i.e., after key exchange) we would expect the entropy to sit close to 8 bits per byte, once encrypted. Shifts in this value might indicate some form of compromise.

- Many legacy protocols still use ASCII encoded plain text encodings. If we detect higher entropy than expected on a known plaintext channel, this may indicate an encrypted channel is being used to send covert messages or exfiltrate sensitive data (e.g., by using encrypted email, or DNS tunnels [HOM17]).

### D. Other Potential Uses of Entropy in Anomaly Detection

In the literature there are studies citing the use of entropy in anomaly detection, and these methods might also be used to characterise and fingerprint a particular infrastructure. For example, entropy can be used to characterise use of IP address, TCP and UDP Port ranges. This may give valuable insights.

For example, we can use the same technique to that describes in Section III to estimate entropy for features such as:

- Packet attributes over time
- IP Addresses and IP Address Pairs
- Port ID and Port ID Pairs
- Timing intervals
- Packet classification
- Flow composition changes across time

The entropy of a set number of attributes with packets can be tracked to assess changes in entropy over time, as described in [ALTH15].

Address and port number entropy (calculated individually or as flow pairs) may give some insights on whether the allocation process for such values appears to be synthetic (or has bugs in the implementation). Entropy in these identifiers may also be used to draw conclusions about the variety of endpoints and services within a packet trace or live network.

Timing (such as packet intervals) can also be a strong indicator of synthetic behaviour. For example, in a denial of service (DOS) attack or brute force password attack, regular packet intervals may be an indicator that the attack is scripted. Even where some randomness has been introduced by the adversary it may be possible to infer higher predictability that expected (for example where a weak random number generator has been used).

Packets may be classified as encrypted on unencrypted using entropy estimates, for example

---

[28] In general encryption tends to produce the highest entropy values compared with compression. Further, naive compression techniques may not achieve high entropy

as described in [DORF11]. This may be problematic if only the first packet payload is used (as in [DORF11]), since early-stage protocol interactions (such as key exchange) may not reflect subsequent higher entropy values.

As discussed earlier, by measuring entropy deviations across the lifecycle of a flow, by flow direction, we may be able to indicate that a flow has been compromised (for example during a masquerade attack, or where a particular encryption method has been subverted [GOUB06]).

Finally, we should keep in mind that skilled malware authors may attempt and hinder entropy-based detection by building synthetic randomness into malware, although it seems promising that weighted or conditional entropy could be deployed across several features to identify outliers.

V. CONCLUSIONS

In this paper we provided baseline payload information entropy metrics across a broad range of common network services, by analysing several widely used datasets in cybersecurity research. To the best of our knowledge this data has not been published previously - at least not comprehensively. From our analysis, mean information entropy values for packet payload are generally consistent across a range of packet capture environments and illustrate the varying degrees of data protection provided by Internet and enterprise services, with subtle differences in inbound and outbound directions. These metrics may be used to approximate *ground truth* for efficiently characterising encapsulated content, from which it should be feasible to help identify certain types of anomalous behaviour. Whilst payload information entropy alone is insufficient to detect broader classes of suspicious behaviour, it can be useful to help identify unusual network behaviour, particularly when correlated with other features, such as flow direction, source and destination network addresses, destination port, timing, state flags, and complementary volumetric features such as payload size and transfer rate.

VI. FURTHER WORK

Since entropy features are rarely published in flow datasets this represents an interesting area from which to perform additional intrusion and outlier detection research, particularly when combined with other features used to classify cyber threats. In future analysis we intend to provide additional fine-grained metrics that further characterise entropy variance deviation and timing changes, by flow direction, during a flow lifecycle, to assist in detecting subtle compromises and man-in-the-middle (MIM) attacks. We also intend to extend the number of datasets analysed.